# Moving the Needle: What Works Cities and the use of data and evidence

Simone Brody
What Works Cities | Results for America
New York, NY, USA
simone@results4america.org

Andel Koester
What Works Cities | Results for America
New York, NY, USA
andel@results4america.org

Zachary Markovits
What Works Cities | Results for America
New York, NY, USA
zach@results4america.org

Jacob Phillips
What Works Cities | Results for America
New York, NY, USA
jacob.phillips@yale.edu

## ABSTRACT

Bloomberg Philanthropies launched What Works Cities (WWC) in 2015 to help cities better leverage data and evidence to drive decision-making and improve residents' lives. Over three years, WWC will work with 100 American cities with populations between 100,000 and 1,000,000 to measure their state of practice and provide targeted technical assistance. This paper uses the data obtained through the WWC discovery process to understand how 67 cities are currently using data to deliver city services.

Our analysis confirms that while cities possess a strong desire to use data and evidence, government leaders are constrained in their ability to apply these practices. We find that a city's stated commitment to using data is the strongest predictor of overall performance and that strong practice in almost any one specific technical area of using data to inform decisions is an indicator of strong practices in other areas. The exception is open data; we find larger cities are more adept at adopting open data policies and programs, independent of their performance using data overall. This paper seeks to develop a deeper understanding of the issues underlying these findings and to continue the conversation on how to best support cities' efforts in this work.

## 1. INTRODUCTION

Bloomberg Philanthropies launched What Works Cities (WWC) in 2015 to help U.S. cities better leverage data and evidence to drive decision-making and improve their residents' lives. Over three years, WWC will work with 100 cities with populations between 100,000 and 1,000,000 to measure their state of practice and provide targeted technical assistance to improve their use of data within city government. This paper uses the data obtained through the WWC discovery process to take a first look at how its cities are currently using data and evidence and see what drives success in cities' practices.

## 2. DATA COLLECTION: THE WWC DISCOVERY PROCESS

Before being selected officially to be a What Works city, cities with populations between 100,000 and 1,000,000 must go through a discovery process. During this process, WWC collects data to better understand each city's current use of data and evidence and, as a result, to inform whether and how we will work with a city to improve its practices.

The WWC discovery process begins when a city submits a statement of interest. The statement of interest is a survey that asks for basic information on the city government and gauges its understanding of the value of data-driven decision-making and its interest in improving this work through WWC. Cities that meet basic qualifying criteria and demonstrate a commitment to the program from their chief executive(s) are invited to complete a self-assessment survey, which provides a more detailed exploration of the city's current data practice. The self-assessment survey is completed separately by at least five city employees, from different departments and roles, each of whom have distinct and cross-cutting views as to how the city uses data and evidence. Concurrently, WWC partners conduct independent research on a city's practices to both validate and expand on the information collected from the surveys.

The WWC discovery process culminates with a visit to the city in which WWC partners engage city leadership and staff in an in-depth discussion on their current practice, goals and challenges in using data to make decisions. Following the visit, WWC makes a final determination about how to move forward with a city based on the data collected throughout the discovery process.

In total, WWC tracks 152 indicators throughout the discovery process, including 89 from the self-assessment, 31 from independent research, 30 from the site visit, and 2 from the statement of interest. Each of these indicators is scored on a 5-point Likert scale that quantifies the city's adherence to the statement it is being evaluated against. This data collection process is fundamental to deriving the measures that we use to evaluate how effectively WWC cities are implementing best practices across data and evidence.

This paper incorporates data from 28 cities that were on boarded through the discovery process described above. Also included are data on an additional 39 cities that entered the initiative through an earlier version of this process, which used a similar, but less rigorous, data collection methodology[i]. Despite this change, the data collected from both groups of cities are similar enough to include in our analysis.

## 3. CONFIRMING THE DATA GAP

In March 2016, WWC released the report, *What Works Cities Brief: The City Hall Data Gap*, which outlines the results of an analysis by Bridgespan Group on data collected from WWC's first 39 WWC cities. This report showed that, while city leaders believe they can foster innovation and solve problems by harnessing the power of data, their cities lack the policies, performance management systems, organizational culture of using data and evidence, and know-how to turn positive intention into results.

Furthermore, it found that the gap between a city's belief in the use of, and its actual ability to use, data in city government is caused by a number of barriers, including lack of staff and financial resources, limited knowledge and expertise, lack of trust in the data currently generated by city systems, old and incompatible systems for data collection and analysis, and challenges in communicating the importance of this work to stakeholders.

Our updated analysis largely confirms the findings from the March 2016 report. Table 1 contains a summary of conclusions from the initial report with updated findings from the most recent set of cities (which have gone through the WWC discovery process since the report's release):

**Table 1 (n = 67 WWC Cities)**

| Observation/Indicator | Finding from original Report (n = 39) | Updated Finding (n = 67) | Change from original Report |
|---|---|---|---|
| City has engaged the public on a strategic goal | 81% | 78% | -3% |
| City publicly communicates their progress on meeting that goal | 19% | 39% | 20% |
| City has invested in a tool or platform to release data to the public | 72% | 74% | 2% |
| City has an established process for regularly releasing data publicly | 18% | 19% | 1% |
| City has a performance management program to track progress towards key goals | 64% | 60% | -4% |
| City has a process in place for analyzing and following up on the information | 30% | 29% | -1% |
| City is committed to using data and evidence to make decisions | 70% | 72% | 2% |
| City modifies existing programs based on the results of program evaluation | 28% | 35% | 7% |
| City has at least 1 strategic goal | 84% | 79% | -5% |
| City has an open data policy to codify and institutionalize how they will share data with the public in the long term | 30% | 29% | -1% |
| City compares relative performance across service providers | 27% | 33% | 6% |
| City Reviews outcomes data during the term of the contract | 27% | 29% | 2% |
| City redirects budget allocations based upon measurable impact | 25% | 30% | 5% |
| City makes at least some contract decisions based on the past performance of vendors | 38% | 42% | 4% |

While we observe only small changes across most data points summarized in the table above, we believe any apparent substantive changes are a function of the updated discovery process and our resulting ability to capture more nuanced data around a city's current practices. We do not believe there is a true deviation in the profile of our cities and that the "city hall data gap" identified in the initial brief remains a deep chasm for our set of cities.

## 4. PERFORMANCE ACROSS WWC STANDARD

To orient the work of helping cities overcome this gap, we created the What Works Cities Standard (hereafter, the Standard) to reflect a set of aspirations and activities that create a strong foundation for the effective use of data and evidence within city government. The Standard is made up of four elements—Commit, Measure, Take Stock, and Act—that build on each other and provide a conceptual framework for how city leaders can advance their data and evidence based practices with government:

**Commit**[ii]: What Works Cities leaders make powerful, public commitments to getting better results for their residents by using data and evidence. High scores on Commit require demonstrated acts that go beyond the passive belief in data and evidence.

**Measure:** What Works Cities leaders use the data and tools at their disposal to measure progress and engage residents along the way.

**Take Stock:** What Works Cities leaders consistently review and reflect to measure progress, learn, and make corrections and improvements.

**Act:** What Works Cities leaders use data and evidence to inform major decisions and take action.

Each of the data points collected throughout the discovery phase fall into one of the four elements of the Standard.

Cities were evaluated based on their total points accumulated across each relevant indicator as a percentage of the maximum allowable score, and the data collected from the second set of 28 cities were the only set of data used to analyze how cities were performing against the Standard. Basic descriptive statistics of the data can be found in Table 2:

**Table 2: Average Scores by Element of the Standard (n = 28 WWC cities)**

| Element of the Standard | Average Score | Standard Deviation |
|---|---|---|
| Commit | 52.3% | 14.0% |
| Measure | 33.4% | 15.2% |
| Take Stock | 42.0% | 15.2% |
| Act | 32.0% | 17.6% |
| Overall | 41.0% | 12.9% |

Since its inception, WWC hypothesized that a leader's commitment to using data and evidence in decision-making is a predictor of a city's success in using data. This belief has been driven by conversations with mayors, city managers, and thought leaders in this work from across the United States, but we have not—until now—had an opportunity to test whether this is empirically valid.

**Table 3: Correlations Among the Elements of the Standard (n=28)**

| Correlations Across Standard | Commit | Measure | Take Stock | Act | Overall Correlation |
|---|---|---|---|---|---|
| %Commit | 1 | | | | **0.82*** |
| %Measure | 0.68* | 1 | | | **0.61*** |
| %TakeStock | 0.75* | 0.51* | 1 | | **0.74*** |
| %Act | 0.64* | 0.46* | 0.7* | 1 | **0.68*** |

*indicates statistical significance at 95% confidence

We found high correlation among all Standard measures, indicating that a city that excels in one element of the Standard is significantly more likely to excel in another, and vice versa (Table 3).

We were also able to explore how predictive a city's "Commit" score is against those of the other elements of the Standard. Interestingly, we found that high scores on Commit were the best predictor of performance across other measures, indicated by a correlation coefficient of .82.

Although this is an initial look at this data and more research needs to be done, the findings seem to suggest that a city's demonstrated commitment to data-driven decision-making is a critical foundation of a successful data practice. When city leaders commit to using data, both in front of city staff and the public, they set a vision and expectation that is necessary, although by no means sufficient, for success.

# 5.PERFORMANCE ACROSS FOCUS AREA

To help bridge the data gap, WWC offers discrete technical assistance from world-class experts that can help a committed city advance its practice in specific focus areas, ultimately helping them progress toward best practice as measured by higher marks in the elements of the Standard.

The WWC Technical Focus areas are sets of technical practices that enable a government to use data and evidence in its everyday work. As with the Standard, WWC assesses each city's practice in each Technical Focus area through indicators based on data collected during the discovery phase. The Technical Focus areas are:

**Open Data:** Data that are made publicly accessible by leadership, public employees, and other stakeholders. By providing transparency around city policy formulation, service delivery, and financial and performance information, cities can empower residents with insights into city service delivery and engender public trust.
**Performance Analytics:** A mechanism by which cities use data to review, manage, and advance strategic outcomes and, ultimately, improve the lives of residents.
**Results-Driven Contracting:** The use of research and data by cities to align procurement with city goals in order to improve the results cities achieve with their contracted dollars.
**Low-Cost Evaluation:** Through the employment of randomized control trials or other evaluation measures, cities can make relatively small, but empirically effective, tweaks to programs, leading to more efficient service delivery.

The indicators used to evaluate cities in this portion of the analysis are a subset of the ones used to evaluate cities across the Standard. The indicators have been reassigned such that a higher score indicates more effective implementation of best practices in the corresponding focus area. Cities were evaluated based on their total points accumulated across each relevant indicator as a percentage of the maximum allowable score. Basic descriptive statistics can be found in Table 4:

**Table 4: Practice Advancement by Focus Area (n=28)**

| Focus Area | Average Score | Standard Deviation |
|---|---|---|
| Open Data | 32.1% | 18.7% |
| Performance Analytics | 49.6% | 15.1% |
| Results-Driven Contracting | 40.2% | 24.6% |
| Low-Cost Evaluation | 30.2% | 19.8% |
| Overall | 41.0% | 12.9% |

Based on our first look at this data, cities score relatively highly on their use of performance analytics to drive their work and are less likely to be using low-cost evaluation methods or implement open data policies and programs.

Table 5 looks at the correlations between these Technical Focus areas:

**Table 5: Correlations among Technical Focus Areas (n=28)**

| Correlations Across Focus Areas | Open Data | Performance Analytics | Results Driven Contract | Low Cost Evaluation | Overall Correlation |
|---|---|---|---|---|---|
| Open Data | 1 | | | | **0.26** |
| Performance Analytics | 0.24 | 1 | | | **0.65*** |
| Results Driven Contract | 0.15 | 0.49* | 1 | | **0.51*** |
| Low Cost Evaluation | 0.15 | 0.72* | 0.33 | 1 | **0.58*** |

* indicates statistical significance at 95% confidence

Overall, there exists a strong positive correlation between a city's score in one focus area and their score in another. These relationships mirror what we saw across city performances in the Standard elements, suggesting that cities that are investing in building their data practice are doing so in a comprehensive way. Another possible explanation is that investment in any one Technical Focus area has spillover effects in helping to bolster practice in other areas, even if there is not discrete investment by the city in improving that area.

Regardless of the explanation, the correlations between a Technical Focus area and a city's overall performance is significantly positive and ranges range from .51 to .65 for each focus area—with one important exception: open data. In fact, the relationship between how a city performs on its open data indicators, and how it performs overall, is weak and statistically insignificant.

## 6. IS OPEN DATA DIFFERENT?

The patterns above suggest that the factors that drive a city to embrace open data may be somewhat different from the factors that drive a city to adopt data- and evidence-based practices in general. One possible factor may be a city's size; an initial analysis in Table 6 suggests that a city's population (p=.01), or alternatively, its budget (p=.007), is a statistically significant driver of its score on our open data indicators, with cities with larger populations and budgets performing better across OD measures. Furthermore, this relationship holds even when disaggregating open data indicators into ones that focus on open data policy, such as whether or not a city has an open data policy, as opposed to open data programs or practice, such as whether a city has conducted a data inventory.[iii][iv]

**Table 6: Data from Selected Regressions (n=28)**

| Independent Variable | Population | | | Budget | | |
|---|---|---|---|---|---|---|
| Dependent Variable | Beta (per 1 million residents) | p-value | $R^2$ | Beta (per $1 billion budget) | p-value | $R^2$ |
| Open Data (Policy and Program) | 0.384* | 0.011* | 0.23 | 0.129* | 0.007* | 0.24 |
| Open Data Policy | 1.72* | 0.015* | 0.21 | 0.55* | 0.016* | 0.2 |
| Open Data Program | 1.55* | 0.027* | 0.17 | 0.48* | 0.034* | 0.16 |
| All Other Indicators | 0.104 | 0.396 | 0.03 | 0.025 | 0.529 | 0.02 |

The correlations between population and budget do not exist between any of the other Technical Focus areas, nor did we observe any other significant relationships using other city characteristics, such as the percentage of individuals with a bachelor's degree, the percentage of individuals living under the poverty line, or the percentage of homes that are vacant within a city.

Although these findings are preliminary, there are a number of possible explanations for these trends in open data. For one, it may be reasonable to infer that larger, wealthier cities with the budget to implement open data portals and the staff to manage and maintain the policies and procedures around those portals would be more likely to score high among open data indicators. Moreover, it is possible that a smaller city that has embraced analytics in its evaluation of specific groups of employees, contracts, or programs may not have the budget or capacity to effectively implement a city-wide open data policy or program. This hypothesis—that larger cities have budgetary advantages to implementing open data—was recently explored by our partners at the Sunlight Foundation in a June 2016 article.[v]

Another possibility is that the relationships within the data are driven by peer effects that exist among cities within population cohorts. The idea that cities emulate peers of a similar size and composition could help explain the clustering of open data proficiency among high-population cities.[vi] In fact, the conceptual framework that underlies this hypothesis is the very one that inspires the WWC initiative: While cities of a certain size are different in a number of ways, they face similar challenges that can be tackled using similar approaches. And sometimes, the peer effects—if that is, in fact, what is driving the trends in open data—are misleading and can cause cities with smaller populations or budgets not to engage on open data efforts where they could be successful. We expect to explore this issue further as we continue to work with individual cities and as we strengthen networks and connections among the community of cities engaging with WWC.

---

[i] http://whatworkscities.bloomberg.org/content/uploads/sites/8/2016/03/WWC_Brief_d.pdf

[ii] High scores on the Commit require demonstrated acts that go beyond the passive belief in data and evidence. A few examples of a commit indicators include: "The city has defined goals," "The city has a strategic document that identifies key citywide priorities and/or goals (i.e., strategic plan, visioning statement, etc.)," and "The city's chief executive publicizes his or her intention to govern with data to city departments."

[iii] Open Data Policy indicators quantify the breadth and sophistication of a city's open data policy including measures such as whether that policy "mandates open and machine readable formats," whether that policy "requires the publication of metadata," and whether that policy "specifies methods for determining the prioritization of data sets for release to the public," etc.

[iv] Open Data Program indicators quantify how effectively and thoroughly a city's open data policy has been implemented, including measures such as whether the city "provides a means for the public to provide feedback on published data," whether the "city's open data website provides a useful search feature" and whether "the city maintains a published comprehensive data inventory,", etc.

[v] http://sunlightfoundation.com/blog/2016/06/29/why-are-some-cities-so-good-at-releasing-open-data-pt-2/

[vi] http://sunlightfoundation.com/blog/2014/10/15/all-five-of-the-largest-u-s-cities-now-have-open-data-policies/